
\input phyzzx

\def\ssstyle{\scriptscriptstyle}
\def\ls#1{_{\lower1.3pt\hbox{$\scriptstyle #1$}}}

\widowpenalty=15

\def\refout{\par\penalty-400\vskip2pc
   \spacecheck\referenceminspace
   \ifreferenceopen \Closeout\referencewrite \referenceopenfalse \fi
   \leftline{\bf REFERENCES}\vskip1pc
   \input \jobname.refs
   }
\unlock
\def\refitem#1{\r@fitem{#1.\enskip}}
\lock

\itemsize=14pt
\refindent=19pt
\parindent=14pt
\abovedisplayskip=3pt plus 5pt
\belowdisplayskip=3pt plus 5pt
\frontpageskip=10pt plus .5fil minus 2pt

\def\linebreak{\unskip\break} 
\def\author#1{\vskip14pt\centerline{#1}}

\tenpoint
\tolerance 500
\looseness 500

\twocols
\hsize=7.5cm   
\hoffset=.35cm
\fullhsize=16cm  
\vsize=20.2cm
\voffset=1cm
\baselineskip=12pt
\parskip=0pt
%

\vbox to 15.75pc{\hsize=\fullhsize
\vskip12pt
\baselineskip=12pt
\line{{\twelvebf Electroweak Baryogenesis:  A Status Report}
\hfill
\vbox{\hbox{SCIPP 94/03}
\hbox{January 1994} }  }
\vskip1pc
\leftline{Michael Dine}
\vskip1pc
\leftline{Santa Cruz Institute for Particle Physics}
\leftline{University of California, Santa Cruz, CA~95064, USA}
\vskip1pc
 There are many reasons to believe that there might
be new sources of CP-violation beyond the phases of the
KM matrix.   One of these is the possibility of electroweak
baryogenesis.  We describe
some recent developments in this subject and make some
comments on the possibility that the minimal standard model
might itself be the origin of the baryon asymmetry.
To appear in the Proceedings of the Third KEK Topical Conference
on CP Violation, its Implications to Particle Physics and
Cosmology, 16--18 November, 1993.
\vskip2pc
}

\leftline{\bf 1.~INTRODUCTION}
\vskip10pt

\REF\sakharov{A.D. Sakharov, JETP Lett. 5 (1967) 24.}
It was A.
Sakharov who first recognized that the origin of the matter-antimatter
asymmetry
is a question which may be addressed scientifically \refmark{\sakharov}.
He noted that there are three conditions necessary
for baryogenesis:
\point  The microscopic laws must violate baryon number
conservation.
\point  CP must be violated.
\point  There must be departures from thermal equilibrium
(there must be an arrow of time).

\vskip2pt
\REF\yoshimura{M. Yoshimura, Phys. Rev. Lett. 41 (1978) 281;
E: Phys. Rev. Lett. 42 (1979) 7461.}
\REF\kolbturner{This subject is presented thoroughly
and pedagogically in E.W. Kolb and M.S. Turner, {\it The Early
Universe}, Addison-Wesley (1990).}
Grand Unified Theories (GUT's)
provided the first appealing framework
in which to satisfy these conditions.  Baryon number
violation is almost a defining property of these theories;
there are typically many new fields and couplings,
permitting much room for CP violation, and the
fact that the expansion rate is very rapid for temperatures
of order the unification scale allows for significant
departures from equilibrium. A substantial
amount of work has been done on GUT
baryogenesis \refmark{\yoshimura,\kolbturner},
and it could well be true that baryons were
produced at some extremely high temperature.
However, many currently popular
ideas in cosmology and in particle physics pose problems
for this possibility.  One difficulty arises in theories of inflation.
In such theories,
the baryon asymmetry must be produced after
inflation.  In GUT baryogenesis, this requires
reheating after inflation to an extremely high
temperature; in practice, this is often difficult
\unskip\linebreak

\vbox to 14.25pc{}

\noindent
to achieve without fine tuning.

\REF\gravitinos{M. Yu Khlopv and A. D. Linde, Phys.~Lett.
B138 (1984)
265; J. Ellis, J.E. Kim and D.V. Nanopoulos, Phys.~Lett.
B145 (1984) 181.}
In addition, particle physics models often
predict the
existence of particles which decay late, producing
entropy which dilutes the baryon number, or which are nearly
stable and must be inflated away.   Among the most problematic
of these is the gravitino of supersymmetric models.
In the most popular schemes for supersymmetry
breaking,
this particle has a mass of $100$'s of GeV.  Yet its width
and cross
sections are of order $G_N$, \ie, extremely tiny.  Thus
if gravitinos were in thermal equilibrium, there are far too
many of them today, and they overclose the universe.
Inflation, if it occurs, sweeps away gravitinos,
just as it sweeps away other debris such as monopoles.
However, if after inflation the universe reheats to a
temperature larger than about
$10^9$~GeV, too many gravitinos will still be produced \refmark{%
\gravitinos}.

\REF\firstfoot{Electroweak baryogenesis
raises the possibility that inflation occurred at temperatures
of order the electroweak scale.  This has been explored in
ref.~[\knox].}
\REF\knox{L. Knox and M. Turner, Phys. Rev. Lett. 70 (1993) 371.}
\REF\secondfoot{An excellent review of the
subject is provided by ref.~[\cknreview].  Many of the points
discussed here are treated at greater length in this review;
a few points are new.}
\REF\cknreview{A.G. Cohen, D.B. Kaplan and A.E. Nelson, UCSD
preprint UCSD-PTH-93-01 (1993) to appear in Annual Reviews
of Nuclear and Particle Science.}
\REF\krs{V.A. Kuzmin, V.A. Rubakov and M.E. Shaposhnikov,
Phys. Lett. B155 (1985) 36.}
\REF\farrar{M.E. Shaposhnikov, Phys. Rev.
Lett. 70 (1993) 2833 and Rutgers preprint RU-93-11.}

A number of ideas have been suggested for dealing with these
problems.  The most elegant is the possibility that the
baryon number is created at the electroweak phase transition,
long (?) after inflation \refmark\firstfoot.
It turns out that all three of Sakharov's conditions can
be satisfied here as well \refmark\secondfoot.
As for condition $1$,
it has been known for many years that the weak
interactions do not conserve baryon number.  While
all vertices of the standard model respect $B$, there
are effective $B$-violating interactions, which
are exponentially small in $\alpha_{\ssstyle W}$.
At low temperatures
and energies, this baryon violation is astronomically small -- in fact,
smaller than that, since no baryon has decayed this way anywhere
in the observable
universe in the age of the universe.
At high
temperatures, however, the rate is enhanced \refmark{\krs};
$$\Gamma = \kappa (\alpha_{\ssstyle W} T)^4.      \eqn\hightemprate$$
This is just want we want in order
to obtain a baryon asymmetry:  the rate is negligible at
low temperatures, but important at high temperatures.

As for the second condition,
CP-violation, of course, exists in the minimal standard
model (MSM).  One expects that this is not  enough to give
the observed asymmetry (see, however, ref.~[\farrar]
and the discussion below), but virtually all popular
extensions of this theory
(multi-Higgs models, supersymmetry, technicolor) offer
the possibility of much stronger CP-violation.
For particle physicists, this close connection with
new physics at weak-scale (LHC, B-factory?) energies
is probably the most interesting
aspect of electroweak baryogenesis.

Finally, there is the question of departure from equilibrium.
The weak phase transition occurs rather late, at a time when
the universe is expanding quite slowly compared to weak interaction
time scales.  (The age of the universe is of order $10^{-8}$ seconds
or so).
It has been known for a long time, however, that if the Higgs mass
is not too large in the minimal model, the electroweak phase
transition is first order (indeed, some of the first inklings of
inflation came from considering phase transitions in theories
with very light Higgs).  A first order phase transition is an
inherently non-equilibrium phenomenon.  It is accompanied
by some amount of supercooling, followed by the formation
of bubbles of the true ground state which grow and fill the
universe.  In most scenarios for electroweak baryogenesis,
it is the surfaces of these bubbles which are the sites of baryon
production.

To really see whether a given model (or any model!)
can produce the observed asymmetry ($n_b/s\sim 0.6-1\times 10^{-10}$)
it is necessary to look at all three of these issues
in more detail.  During the last few years, there has been
a great deal of work on each of them.
This we will review in the subsequent sections.

\vskip2pc
\leftline{\bf 2.~BARYON NUMBER VIOLATION IN}
\leftline{\bf THE ELECTROWEAK THEORY}
\vskip1pc
Consider, first,
the question of baryon number violation.  At first sight,
baryon number violation in the standard model is a bit
of a surprise, since all of the Feynman diagrams of the
theory conserved $B$ (indeed, this is one of the triumphs
of the standard model, and one of the puzzles which must
be explained in extensions of the theory such as supersymmetry).

But this situation can be understood in rather simple terms.
Baryon number violation in the standard model arises
because the baryon number current is anomalous:
$$\partial_{\mu}j^{\mu}_B = {3g^2 \over 32 \pi^2} F^a \tilde F^a.
\eqn\baryonanomaly$$
The right hand side of this equation turns out to be a total
divergence, $\partial_{\mu}K^{\mu}$, where $K^{\mu}$ is
known as the Chern-Simons current.
Integrating both sides of this equation over space-time,  we
see that in any process
$\Delta B = \Delta n_{cs}$, where $\Delta n_{cs}$, is the
change in the
``Chern-Simons number," essentially the ``charge" associated
with $K^o$.

\REF\manton{N.S. Manton, Phys. Rev. D28 (1983) 2019;
F.R. Klinkhammer and N.S. Manton, Phys. Rev. D30 (1984)
2212.}
The significance of this equation becomes clear when one examines
the vacuum structure of non-Abelian gauge theories.
In these theories,
the classical, zero energy configurations (in suitable gauges)
are labeled precisely by $n_{cs}$, which can
take any integer value.   In
strongly coupled theories like QCD, the implications of this
statement are obscure, but in weakly coupled theories
like the standard model in its broken phase,
this means that there are a set
of classically degenerate states labeled by $n_{cs}$.  Quantum
mechanically, these states differ in baryon number (and
as a result carry slightly different energy).  These states
are separated from one another by a barrier.  As a result, at
low temperatures, it is necessary to tunnel from one state
to another, and the rates are exponentially small, behaving
as $e^{-2 \pi /\alpha\ls{\ssstyle W}}$.  At high
temperatures, however,
thermal fluctuations can take one over the barrier, and the
rate is simply governed by a Boltzmann factor, $e^{-\beta E_B}$,
where $E_B$ is the barrier height.  In a field theory, the
configuration at the top of the barrier corresponds to a static solution
of the classical equations of motion with a single negative mode,
known as a ``sphaleron" \refmark{\manton}.
Simple scaling arguments give for the energy of this configuration,
$E_B \sim M_{\ssstyle W}/\alpha_{\ssstyle W}$.

\REF\shaposhmc{J. Ambjorn, T. Askgaard, H. Porter and
M.E. Shaposhnikov, Phys. Lett. B244 (1990) 479;
Nucl. Phys. B353  (1991) 346.}
So the rate grows rapidly as the temperature increases.
In fact, $M_{\ssstyle W}$ is itself a function of temperature, which
decreases with temperature, tending to zero at temperatures
of order 100~GeV--1~TeV.   At these temperatures,
this calculation can't really make sense; there
is no small parameter to justify a semiclassical treatment.
Indeed, at very high
temperatures, one does not know how to do a reliable calculation,
except, possibly, within the framework of
lattice gauge theory \refmark{\shaposhmc}.
General arguments, however, show that the rate for
$B$-violating transitions (per unit volume per unit time) goes as
$${\Gamma = \kappa \alpha_{\ssstyle W}^4 T^4.}$$
As we will see, the rates in both the broken and unbroken
phases are important to the
question of baryogenesis.

\REF\carson{L. Carson, X. Li, L. McLerran and R.-T Wang, Phys.
Rev. D42 (1990) 2127.}
\REF\dortmund{J. Baacke and S. Junker, DO-TH-93/15.}
\REF\subtleties{E. Farhi, V.V. Khoze, R. Singleton,
Phys. Rev. D47 (1993) 5551.}
Let us consider, first, the question of computing the rate
at temperatures (well) below the phase transition
temperature greater detail.  In this regime,
it is possible
to do a systematic, semiclassical computation.
In addition to the overall $e^{-E_{\scriptstyle sp}/T}$ factor,
the prefactor can be calculated provided one
performs a (in principal straightforward but)
difficult evaluation of a functional determinant.
Such a calculation was done a few years ago by Carson
\etal~\refmark{\carson}. Recently, however,
Baacke and Junker \refmark{\dortmund} have
performed
a new calculation of the required determinant, and found
a much larger answer.  To the best of my knowledge,
the discrepancy has not yet been resolved.
At finite temperature, there are
also issues of principle in deciding precisely what quantity
to evaluate \refmark{\subtleties}.
Typically, one uses
the finite-temperature effective action to determine the form
of the sphaleron, but this introduces a certain amount
of double counting (as first noted by Carson \etal).

\REF\arnoldm{P. Arnold and L. McLerran, Phys. Rev.
D36 (1987) 581; Phys. Rev. D37 (1988) 1020.}
\REF\vancouver{M. Dine,  in
The Vancouver Meeting:  Particles and Fields, the Proceedings of
the Meeting of the DPF Division of the APS, Vancouver, B.C.,
August 18--22, 1991, D. Axen and D. Bryman, eds.,
World Scientific, Singapore, 1992, p.~831.}
Our knowledge of the high temperature rate is
even more shaky.   The general arguments actually only show
that the rate
goes as $g^8$.  It is simply as a result of prejudice that one
sticks in the $4 \pi$'s, and refers to $\kappa$ as a ``coefficient
of order unity."  There are some lattice calculations which suggest
that indeed $\kappa$ is of order unity \refmark{\shaposhmc}.
However, the analysis
is quite difficult, and the results are very noisy.  (It is difficult,
for example, to see the required diffusive behavior of $n_{cs}$
with time.)  One can also make crude guesses, based on
simplified models of the high temperature rate and extrapolation
from the low temperature rate \refmark{\arnoldm,\vancouver}.
These could give quite large
values of $\kappa$, if the Dortmund results are correct.  These
crude estimates scale linearly with this prefactor.

\vskip2pc
\leftline{\bf 3.~THE ELECTROWEAK PHASE}
\leftline{\bf TRANSITION}
\vskip1pc

In order to study the nature of the electroweak phase
transition, the simplest thing to do is to compute the
free energy for a given, fixed, background Higgs field.
In the MSM, this calculation is easy.  To lowest order, one
just computes the contribution to the free energy
from each particle species, with a mass appropriate to
$\VEV{\phi}$.  In this way, one obtains \ref{Here we ignore
the Higgs contribution.  This is justified provided that
we are considering temperatures for which the effective
Higgs mass is not small, or Higgs which are light compared
to the gauge bosons.  Obviously, this last condition is not
likely to hold in the real world.}
$$
V(\phi,T) = D (T^2 - T_o^2) \phi^2 - E T \phi^3 +
{\lambda_T\over 4} \phi^4 \ .
\eqn\hightemp$$
Here
$$
D = {1\over 8v_o^2} ( 2 m_W^2 + m_Z^2 + 2 m_t^2) \ ,
\eqn\dequals  $$
$$
E = {1\over 4\pi v_o^3} ( 2 m_W^3 + m_Z^3) \sim 10^{-2} \ ,
\eqn\eequals $$
$$
T^2_o = {1\over 2D}(\mu^2 - 4Bv_o^2) =
{1\over 4D}(m_H^2 - 8Bv_o^2) \ ,
\eqn\tzero$$
$$\eqalign{%
\lambda_T = \lambda - {3\over 16 \pi^2 v_o^4}
 &\left(  2 m_W^4 \ln{m^2_W\over a_B T^2}
 + m_Z^4 \ln{m^2_Z\over a_B T^2}\right.  \cr
&~~\left. - 4 m_t^4 \ln{m^2_t\over a_F T^2}\right) \ ,
\cr }
 \eqn\lambdat
$$
where $\ln a_B = 2 \ln 4\pi - 2\gamma \simeq 3.91$,
$\ln  a_F = 2 \ln \pi - 2\gamma \simeq 1.14$.
This expression gives rise to a first order phase transition,
at least for sufficiently small Higgs mass, on account of the
cubic term.  At very high temperatures, the potential
has only a minimum at the origin.  As the temperature
is lowered, a second, local minimum appears.  At a slightly lower
temperature, this minimum becomes the global minimum,
 and it finally disappears at temperatures below $T_o$.

\REF\shaposhnikovlimit{M.E. Shaposhnikov, Nucl. Phys. B387 (1987)
757; A.I. Bockharev and M.E. Shaposhnikov, Mod. Phys. Lett. 2A
(1987) 417.}
Before turning to the question of whether the
phase transition is sufficiently strongly first order
to produce baryons, we can, following
Shaposhnikov \refmark{\shaposhnikovlimit},
establish a criterion which the transition
must satisfy if there is to be any possibility
of generating the asymmetry.  Suppose, somehow, an asymmetry is produced
during the phase transition.  Before the transition, the expectation
value of the Higgs field(s) vanishes; afterwards, it (they) take
some non-zero value, $\phi_o$.  Thus after the phase transition,
the baryon violation rate goes as
$${e^{-E_{\scriptstyle sp}/T},}$$
where $E_{sp}$ is roughly linear in $\phi_o$.
Thus if $\phi_o$ is too small, the $B$-violating rate is large,
and any baryons produced during the transition will
be lost.

\REF\shaposhnikovlatest{M. Shaposhnikov, CERN preprint
CERN-TH.6918/93 (1993); K. Kajantie, K. Rummukainen and
M. Shaposhnikov, CERN preprint CERN-TH.6901/93 (1993).}
\REF\dhll{M. Dine, P. Huet,
R. Leigh and A. Linde, Phys. Lett. B283 (1992) 319;
Phys. Rev. D 46 (1992) 550.}
This argument, plus the present limits on the Higgs mass from
LEP, can be used to pretty well rule out the minimal standard
model as the origin of the baryon asymmetry (see, however,
ref.~[\shaposhnikovlatest]).  Using the lowest order potential discussed
above, one can show that for a Higgs mass above
about 45~GeV, $\phi_o$ is too small.
Taking account of certain one loop corrections to the
Higgs potential, the limit is even stronger \refmark{\dhll}.
Essentially, the coefficient of the cubic term
in eq.~\hightemp\ is reduced by about $1/3$; this makes
$\phi_o$ smaller, and the Higgs limit becomes about
35~GeV.

\REF\turoketal{A.I. Bockharev, S.V. Kuzmin, M.E. Shaposhnikov,
Phys.~Lett. 244B (1990) 27;
N. Turok and J. Zadrozny, Nucl.~Phys. B369  (1992) 729.}
\REF\hallanderson{G. Anderson and L. Hall, Phys.~Rev. D45 (1992) 2685.}
\REF\myint{S. Myint, Phys. Lett. B287 (1992) 325.}
\REF\zwirner{J.R. Espinosa, M. Quiros and F. Zwirner,
Phys. Lett. B307 (1993) 106.}
Models with more structure than that of the minimal standard
model allow more strongly first order phase transitions
with more massive Higgs fields \refmark{\turoketal}.
Anderson and Hall \refmark{\hallanderson}
have pointed out that this goal can be achieved very easily
in models with additional singlets.  The minimal supersymmetric
standard model has been explored in refs.~[\myint] and [\zwirner].
The latter analysis is particularly thorough and up to date.
These authors find that the bounds are somewhat relaxed,
allowing for baryogenesis in a small range of parameter space.

\REF\lindebubbles{A.D. Linde, Phys.~Lett. B70 (1977) 306;
100B (1981) 37; Nucl.~Phys. B216 (1983) 421.}
\REF\mtl{B.-H. Liu, L. McLerran and N. Turok,
Phys. Rev. D46 (1992) 2668.}
\REF\freese{M. Kamiankoski and K. Freese, Phys. Rev. Lett. 69
(1992) 2743.}
\REF\huetetal{P. Huet, K. Kajantie, R.G. Leigh and B.-H. Liu,
Phys. Rev. D48 (1992) 2477.}
In order to understand the origin of any possible
asymmetry, it is important to understand the
dynamics of the phase transition.
Given that the transition in a particular model is reasonably
strongly first order, there is a standard
theory of how bubbles of ``true
ground state" form \refmark{\lindebubbles},
but the question of how they propagate
in the finite temperature plasma has turned out to be a rather
difficult one.  For the problem of baryogenesis,
the most important parameters are the bubble velocity
and the thickness (asymptotically) of the wall.
After some initial controversy, a rough
consensus has emerged on how to determine these parameters.
For example, for a Higgs (unrealistically) light enough
in the minimal standard model to give a first order transition,
one finds typical velocities of order 0.1--0.4~c and thicknesses
of order 10--20 T$^{-1}$ \refmark{\dhll,\mtl}.
In more complicated models with
more strongly first order transitions, thinner and faster walls
arise.
There has also been some dispute over whether the bubble
walls are nice, smooth surfaces, or whether they are more dendritic
in character \refmark{\freese,\huetetal},

\REF\arnold{P. Arnold and O. Espinosa, Phys. Rev. D47 (1993) 3546.}
\REF\bagnasco{M. Dine and J. Bagnasco, Phys. Lett. B303 (1993) 308.
\endpage}
\REF\kg{M. Gleiser and E.W. Kolb, Phys. Rev. Lett. 69 (1992) 1304.}
\REF\anderson{G. Anderson, Phys. Lett. B295 (1992) 32.}
\REF\tetradis{N. Tetradis, Z. Phys. C57 (193) 331.}
\REF\marche{J. Marche-Russell, LBL preprint LBL 34573 (1993)
and Phys. Lett. B296 (1992) 364.}
\REF\ay{P. Arnold and L. Yaffe, Washington preprint UW/PT-93-24 (1993).}
Even in the calculation and interpretation of the
effective potential, there
are many subtleties, and this
has led to a good deal of controversy in the literature.
The question basically turns on what is the small parameter,
and is it small enough in cases of practical interest.
A particularly thorough analysis, from this perspective,
has been performed recently by
Arnold and Espinosa \refmark{\arnold},
who computed all of the relevant diagrams to two-loop
order, and considered their numerical values (a simplified
version of this analysis, picking out certain leading contributions,
was performed by J. Bagnasco and myself \refmark{\bagnasco}).
The result is that
away from the phase transition, perturbation theory
is not so bad (corrections are typically of order $10\%$
for a 35 ~GeV Higgs).
Near the transition, corrections are larger, but the limits
on the Higgs mass do not change appreciably.  Still, one
expects that as the Higgs mass increases, perturbation theory
is not a reliable guide.  An optimist might
hope that as the Higgs mass increases, even though
perturbation theory suggests that the transition is more and more
weakly first order, perhaps this is not correct; perhaps the
transition is still first order.  A number of analyses have been put
forward which
suggest that this is not the case.  First, Kolb and Gleiser \refmark{\kg}
have argued
that as the Higgs mass is increased, a picture of distinct phases
coexisting near the transition point is misleading, because
fluctuations are so large.  To describe this situation in a quantitative
way, they have developed a theory of ``subcritical bubbles."
Other analyses of the transition, particularly for higher
mass Higgs, have been performed by Anderson \refmark{\anderson},
Tetradis \refmark{\tetradis},
Marche-Russell \refmark{\marche}, and others.  A particularly
thorough analysis has just recently appeared by Arnold and
Yaffe \refmark{\ay}. These authors compare systematically the results
of the $\epsilon$ expansion with those of conventional perturbation
theory, both for thermodynamic quantities such as the free energy and for the
sphaleron rate.  While their results suggest that the phase transition
in the minimal model is more first order than indicated by perturbation
theory, they also suggest that sphaleron transitions are more rapid.

The more optimistic viewpoint has been advocated
recently by Shaposhnikov and collaborators, both
on the basis of theoretical arguments and lattice
simulations \refmark{\shaposhnikovlatest}.
These authors argue that even for rather small Higgs mass,
perturbation theory is completely misleading and that
the transition is strongly first order
even for rather large
Higgs mass.  I am personally
rather skeptical of these analyses, but there are certainly
good reasons to be suspicious of perturbation theory, and
improved simulations are clearly important.  Even comparisons
of perturbation theory and lattice results for pure gauge theory
would be of some interest in this regard.  In particular,
the arguments of ref.~[\shaposhnikovlatest] imply that there
should be significant discrepancies here.

\vskip2pc
\leftline{\bf 4.~BARYOGENESIS}
\vskip1pc
\REF\antifarrar{M.B. Gavela, P. Hernandez, J. Orloff
and O. Pene, Harvard preprint HUTP-93/A036 (1993).}
Finally we turn to the problem of the actual
asymmetry
produced in different models.  The most obvious question, perhaps,
is could the minimal standard model itself be the origin
of the baryon asymmetry?  The conventional wisdom is no:
first, as noted above, the standard model Higgs is too massive,
and second, there is too little CP-violation; since
all three generations are involved, one expects
suppression
by mixing angles and quark Yukawa couplings.  Recently,
however, Farrar and Shaposhnikov have discussed a scenario
in which they claim one can obtain a quite hefty asymmetry.
I am, again,
rather skeptical about this claim (it has been discussed
at some length by Shaposhnikov at this meeting).  I will comment
about
it to some extent below, after having discussed baryogenesis
in various extensions of the MSM.  We will see that in a certain
limit -- which is likely to be the physically interesting one --
the baryon asymmetry is straightforward to calculate, and it
is extremely tiny, containing all of the expected suppression
factors and more!  The authors of ref.~[\farrar] assume a different
limit, but even here it seems unlikely that there is any
appreciable asymmetry.  Their estimate relies on
very optimistic assumptions about rates, mean free
paths, and properties of the transition. Recently a preprint has
appeared arguing that
the delicate coherence required for their effect
does not exist in the high temperature plasma,
and that there is virtually no asymmetry produced,
even granted their other assumptions \refmark{\antifarrar}.

While it would be extremely economical if the minimal model
could produce the observed asymmetry, some might view it
as more exciting if only extensions of the minimal theory
could produce the asymmetry.  One would then have the
hope of making new predictions for particle physics,
provided one assumed baryon production at the electroweak transition.
Virtually all extensions of the minimal theory possess
new sources of CP-violation.  Many of these theories
appear capable of producing the observed asymmetry.

Most of the literature has focused on two types
of scenarios for generating the baryon asymmetry.
The first assumes that the
bubble wall is slowly moving, and rather thick.  Then
the Higgs field changes slowly near the surface of the
bubble, and the transition is ``adiabatic," in the sense that
virtually everything remains in equilibrium except the
baryon number.  The alternative is that the wall moves rapidly
and is thin.

\REF\dineetal{M. Dine, P. Huet, R. Singleton and L. Susskind,
Phys. Lett. 256B (1991) 351.}
\REF\mstv{L. McLerran, M. Shaposhnikov, N. Turok
and M. Voloshin, Phys. Lett. B256 (1991) 451.}
\REF\ckntwo{A.G. Cohen, D.B. Kaplan and A.E. Nelson, Phys. Lett.
B263 (1991) 86.}
\REF\withscott{M. Dine and Scott Thomas, SCIPP preprint 94/01 (1994)
submitted to Phys.~Lett.~B.}
The adiabatic case is the easiest
to analyze (though it was not studied first).  It
was first considered in refs.~[\dineetal], [\mstv]
and [\ckntwo].  Here I would like to give a description
which is an extension of the analyses of ref.~[\dineetal]
and [\mstv], but I believe also somewhat simpler and
more correct.  The ideas I describe here were developed
in collaboration with Scott Thomas, and appear in ref.~[\withscott].
In this regime, baryon number is produced because the
time-varying Higgs field in the bubble walls
biases the sphaleron process, so
that production of baryons is preferred over baryons.
There are two basic ingredients to this analysis.
The first is the observation that the baryons are produced
only for a very short time, $t_{co}$.  This is because
the sphaleron process shuts off rapidly as the Higgs
field turns on; recall that the sphaleron rate goes as
$e^{-Ag\phi/\alpha_{\ssstyle W} T}$, where $A$ is a
number of order unity, so the process cuts off
as soon as $g \phi_{co} \sim \alpha_{\ssstyle W} T$.  Simple estimates
of reaction rates show that at $t_{co}$
the densities of the different particle species have hardly
changed from their original values.
The second ingredient
is the rate equation.
The basic idea is that if two states of different $n_{cs}$
differ in free energy by an amount $\mu$, then by
detailed balance, the rates for transitions which increase
$n_{cs}$, $\Gamma_+$, and those which decrease $n_{cs}$,
$\Gamma_-$, are related by
$${\Gamma_+ \over \Gamma_-}=e^{-\beta \mu}.\eqn\detailedbal$$
If $\Gamma$ is the average rate, than the rate of change of baryon
number is given by
$${dn_B \over dt}=3 \Gamma  \mu/T.\eqn\baryonrate$$

\FIG\trianglegraph{Leading coupling of the Higgs phase
to the Chern-Simons number in the two Higgs doublet model.}
\REF\finitet{See, for example, S.A. Abel, W.N. Cottingham
and I.B. Whittingham, Rutherford preprint RAL 93-2.}
For definiteness, let us consider a particular model, the
two Higgs model discussed in refs.~[\mstv] and [\ckntwo].
In this model, it is the coupling of one of the Higgs fields,
$\phi_1$, to top quarks which gives the largest effects.
Call this coupling $\lambda_t \phi_1 \bar t t + h.c.$, and
write $\phi_1=\rho_1 e^{i \theta_1(x,t)}$.  We want to understand
how the time-varying phase, $\theta_1$, couples to
the Chern-Simons number.  The most naive thing to do
is simply to compute the triangle graph in fig.~\trianglegraph.
A straightforward calculation, using the rules of finite temperature
perturbation theory, gives for this coupling \refmark{\finitet}
$${\cal L}_{\theta}= a \theta_1{ m_t^2 \over  T^2}
{F \tilde F \over 32 \pi^2}+ {\cal O}\left({m_t\over T}\right)^4
\eqn\thetaffcoupling$$
where $a={14 \over 3 \pi^2} \zeta(3)$.
Integrating by parts gives
$$-a{m_t^2 \over T^2} \partial_o \theta_1
 n_{cs}.\eqn\cscoupling$$
If $\theta_1$ is sufficiently slowly varying, this is, indeed,
a chemical potential for Chern-Simons number.  As noted above,
we can ignore any tiny CP-violating densities which might
develop before $t_{co}$, so eq.~\cscoupling\ represents the largest
contribution to the free energy difference of the states.
We can plug this into our rate equation above, integrate,
and proceed to estimate the asymmetry.  In the adiabatic
limit, this calculation is simplified because we can take the
instantaneous rate to be that associated with the instantaneous
value of $\phi$.  Recalling that the process
shuts off for $\phi = \phi_{co}$, and that $m_t
\propto \phi$, the overall
asymmetry is of order
$${n_B \over s}\sim 10^{-2} \Delta \theta_1 \phi_{co}^2 \kappa
\alpha_{\ssstyle W}^4.\eqn\asymmest$$
In the multi-Higgs model, $\Delta \theta_1$ is
itself of order $\phi_{co}^2$,
so the asymmetry is potentially {\it quite} small.

\topinsert
\vskip-10.5cm
\vskip-7cm
\vskip6pt\noindent
Figure \trianglegraph.\enskip
{Leading coupling of the Higgs phase
to the Chern-Simons number in the two Higgs doublet model.}
\vskip1pc
\endinsert

The actual value of $\phi_{co}$ is a subject of
controversy.  In the broken phase, the sphaleron
rate goes as
$$\Gamma(\phi) \sim M_w^7(\phi) e^{-{A \beta M\ls{\ssstyle W}}(\phi)/
\alpha\ls{\ssstyle W}}.\eqn\largephirate$$
This formula exhibits a peak for $M_{\ssstyle W}=7 \alpha_{\ssstyle W} T$.
This fact has been used to argue that the suppression is not
so large.  However, this formula does not exhibit the
correct behavior for small~$\phi$.  In ref.~[\vancouver],
a toy model was written down which has the correct behavior
in both the large and small $\phi$ limits, and $\phi_{co}$
turns out to be two to three times smaller.  On the
other hand, this model also suggests that $\kappa$ may be
quite large.  Of course,
all of these arguments are heuristic.
These questions must ultimately be settled
by improved simulations.  Clearly we need the effective
$\kappa$ to be quite large, of order $10^{3}$ or larger,
if we are to obtain an adequate asymmetry in this model.

The extension of this calculation to other theories is
straightforward.   For example, in the MSSM, one does
a little better, in that one does not obtain the factor
of $\Delta \theta^2$.  On the other hand, one pays the
price of some explicit CP-violating phases, which
are expected to be small (see below, however).

\REF\fourthfoot{In particular
cases, this transformation has been criticized in the literature,
since the transformation involves a gauge symmetry, and
gauge charges are screened \refmark\khlebnikov.
However, this problem appears to be
a red herring; in practice, one can perform
\unskip\linebreak
\endpage
\itemsize=19pt
\item{}
a transformation involving non-gauged currents,
for which there is no problem of screening \refmark{\cknreply}.}
\REF\khlebnikov{S.Y. Khlebnikov, Phys. Lett. B300 (1993) 376.}
\REF\cknreply{
A.G. Cohen, D.B. Kaplan and A.E. Nelson, Phys. Lett. B294 (1992) 57.}
\REF\fifthfoot{Shaposhnikov and
Giudice have recently pointed out that if QCD sphaleron processes
are rapid, that the minimum of the free
energy lies at
zero baryon number, up to terms of order quark
mass-squared \refmark\guidice.}
\REF\guidice{G. Giudice and M.S. Shaposhnikov,
CERN preprint CERN-TH-1080-93 (1993).}

An alternative treatment of the two-Higgs model
has been suggested by
Cohen, Kaplan and Nelson \refmark{\ckntwo}.
The basic idea is to write the Higgs fields as
$${\phi_i(t)=\rho_i(t)e^{i \theta_i(t)}}$$
and to eliminate the phases by suitable transformations of the
Higgs fields and fermions.  In the case that all of the quark
masses arise from a single Higgs field, this can be
achieved by a hypercharge rotation of the fermions \refmark{\fourthfoot},
$$\psi_i \rightarrow e^{-iy_i \theta_1}.\eqn\hypercharge$$
It is convenient to parameterize the Higgs field as
$$\phi_1=e^{i \theta_1} (\rho_1 + \phi^{\prime})
\eqn\phiparameterization$$
where $\phi^{\prime}$ represents the fluctuating part
of the Higgs field.
This eliminates the phases from the Yukawa couplings,
but leaves an effective interaction of the form
interaction of the form \refmark\fourthfoot
$${\dot \theta_1 \rho\ls Y}$$
where $\rho\ls Y$ is the hypercharge density (including the fluctuating
scalar field).
If $\theta_1$ is varying adiabatically in the sense described earlier,
this is like a chemical potential for hypercharge;
the resulting minimum of the  free
energy lies at non-zero baryon number \refmark\fifthfoot.
Since baryon number
is violated, one expects that
the system tries to get there.

However, the chemical potential here is only relevant for a very
short time, the time it takes for the baryon number violating
process to shut off.  All that interests us is the rate equation
valid during this short time interval.  To find it, we can
again ignore
everything but the baryon-number violating sphaleron process,
and take all
CP-violating densities to be approximately zero.
Then if we ignore the small quark Yukawa
couplings, fermionic hypercharge
is conserved.  In particular,
the sphaleron process respects it, so there is no
biasing!  To obtain any effect,
one must take account of the Higgs VEV, \ie, of the
effective fermion masses.
In the presence of the mass
term, the divergence of the current is no longer zero; it
has an additional piece, which is precisely the coupling
of eq.~\thetaffcoupling\ above.  This can be interpreted
as meaning that, on average, there is a change in the hypercharge
current in a sphaleron transition.  Taking this into account,
one obtains
precisely the result of the naive treatment above \refmark{\withscott}.

\REF\cknone{A.G. Cohen, D.B. Kaplan and A.E. Nelson,
Phys. Lett. B245 (1990) 561; Nucl. Phys. B349 (1991) 727.}
\REF\cknthree{A.G. Cohen, D.B. Kaplan and A.E. Nelson, Nucl. Phys.
B373 (1992) 453.}
A more substantial asymmetry may arise in the opposite limit,
where the wall is thin and rapidly moving.
This point was first stressed by Cohen, Kaplan and Nelson \refmark{%
\cknone,\cknthree}.
For example, in a model with several Higgs doublets, the principle
contribution arises from scattering of top quarks off of the
bubble wall.  Because CP is violated, an asymmetry can develop
between the number of left-handed and right-handed top quarks
in the region in front of the wall.  This asymmetry, in turn, biases
the rate of baryon number production in this region.  This mechanism
is particularly efficient because the departures from equilibrium
are large, and also because the baryon density is produced in a region
with vanishing Higgs field, so the suppression discussed above is absent.
According to ref.~[\cknthree], the baryon to entropy ratio in this
case can be as large as $10^{-4}$ or so, times CP-violating phases.
(This is actually the sort of mechanism contemplated by
Shaposhnikov and Farrar \refmark{\farrar}.)

\vskip2pc
\leftline{\bf 5.~RECENT DEVELOPMENTS, AND}
\leftline{\bf SOME CONCLUSIONS}
\vskip1pc

\REF\comelli{D. Comelli, M. Pietroni and A. Riotto, DFPD 93/TH/26.}
\REF\schmidt{M. Peskin and C.R. Schmidt, Phys. Rev. Lett. 69 (1992)
410; C.R. Schmidt, Phys. Lett. B293 (1992) 111.}
\REF\gordy{C.J.C. Im, G.L. Kane and P.J. Malde, University of Michigan preprint
UM-TH-92-27 (1992).}
\REF\gunion{B. Grzadkowski and J.F. Gunion, Phys. Lett. B294
(1992) 361.}
Over the past year or so, there have been a number of further developments.

One which I find particularly appealing is due to Comelli
\etal~\refmark{\comelli}.
These authors consider a variant on these ideas, in which explicit
CP-violation is small, as in supersymmetric models, but CP is
{\it spontaneously} violated at the phase transition.  One's first thought
is that this can't help, since in the absence of the small, explicit
CP-violation, the spontaneous violation will just lead to equal numbers
of domains with one sign or the other of the baryon number.  But
what these authors point out is that the small CP-violation is enough
to give a significant preference to domains of one type or the other.
The point is that the rate for tunneling is the exponential of a large,
negative number; small fractional changes in this number lead to large
numerical changes in the rate.

Farrar and Shaposhnikov
have argued that the MSM might be the origin of the
observed asymmetry \refmark{\farrar}. We have already mentioned
some criticisms which have been leveled at this idea.
It is amusing to calculate the asymmetry in the
adiabatic limit, along the lines we have just described.
(This is the opposite limit to that assumed by these
authors.  Perturbation theory gives a transition which
looks close to the adiabatic limit, but these authors have
also argued that perturbation theory is not a reliable
guide to the nature of the transition.)   One again just
computes the leading coupling of the time-varying Higgs
field to the Chern-Simons number.  One now finds an
operator
$$\vert \phi \vert^2{g^2 \over 16 \pi^2} F \tilde F \sim
\partial_o \vert \phi \vert^2 n_{cs}.\eqn\standardmodel$$
This operator cannot be generated before $7$ loop order.
It is suppressed by quark masses and mixing angles (giving
a factor of order $10^{-20}$), as well as two powers of $\phi_{co}$,
four powers of $\alpha_{\ssstyle W}$, and loop factors which are not likely
to be larger than $10^{-10}$.  So in this limit, at least, the asymmetry
is ridiculously small.

Finally, what are the implications of all of this for particle physics?
Let me suppose, for a moment, that we have reached a stage at which,
for a given model, we can reliably compute the resulting asymmetry.
If, at the SSC or elsewhere, one finds evidence for new physics, one
can ask whether this physics (or what values of the parameters
associated with this physics) could produce the observed asymmetry.
For example, if these ideas are correct, one expects that CP-violation
is likely to be substantial.  Could one see it?  Motivated in part by these
considerations, there has been some effort to determine how such
a CP-violation could be measured.  C. Schmidt and
M. Peskin have examined the question of whether one could see an
asymmetry in the energies of leptons and antileptons in top decays at hadron
colliders
\refmark{\schmidt}.
They define a CP-violating asymmetry in top quark decay
that is difficult \refmark{\gordy},
but perhaps not hopeless, to measure.
Gunion and collaborators
have discussed the prospects for observing CP-violation
in the Higgs sector at the NLC and hadron
colliders \refmark{\gunion}.
Such CP-violation
is reflected in the fact that the Higgs mass eigenstates
don't have definite CP-properties.
Higgs sector CP-violation
is not likely to be important in the framework of the simplest
supersymmetric models, but it can arise in multi-Higgs theories
and in non-minimal supersymmetric theories.
The  NLC suggestion involves using the machine as a photon-photon collider,
and looking at the couplings of the photons to the different mass
eigenstates.  At hadron colliders,
one uses polarized proton beams, and considers
the analogous effect involving gluons.

In conclusion, it appears quite plausible that the
baryon asymmetry arose at the electroweak phase
transition.   A number of mechanisms for producing the
asymmetry have been suggested.  Real quantitative
progress, however, will require much more extensive
simulations both
of features of the phase transition and of baryon
violating rates.

\refout
\bye